\documentclass{ws-procs9x6}

\begin{document}

\title{Neutrino Physics with the IceCube Detector}

\author{J. Kiryluk for the IceCube Collaboration}

\address{
Lawrence Berkeley National Laboratory, \\
1 Cyclotron Rd, Berkeley, CA, 94720, USA \\
E-mail: JKiryluk@lbl.gov}  

\maketitle

\abstracts{
IceCube is a cubic kilometer neutrino telescope under construction at the South Pole.
The primary goal is to discover astrophysical sources of high energy neutrinos. We 
%discuss the IceCube neutrino physics program,
describe the detector 
and present results on atmospheric muon neutrinos from 2006 data collected with
nine detector strings.}

\section{Introduction}

The IceCube detector is a cubic kilometer  neutrino telescope under construction at the South Pole\cite{icecube}. The main goal of IceCube is to detect cosmic neutrinos of all flavors 
in a wide (100 GeV to 100 EeV) energy range\footnote{So far the only observed extra-terrestrial neutrinos are low energy neutrinos from the Sun\cite{Sun} and SN1987A\cite{SN}.}.
IceCube will search for  point sources of extra-terrestrial muon neutrinos 
and diffuse fluxes of extra-terrestrial neutrinos of all flavors. 
Possible high energy neutrino sources are active galactive nuclei (AGNs),
gamma-ray bursters (GRBs) and supernova remnants (SNRs). 
Other IceCube physics topics include searches for 
WIMP anihilation in the Earth and Sun, signatures of supersymmetry in neutrino interactions, and exotica like magnetic monopoles or extra dimensions\cite{icecube}.  

\section{The IceCube detector}

\begin{figure}[t]
\begin{center}
\includegraphics[width=0.8\textwidth]{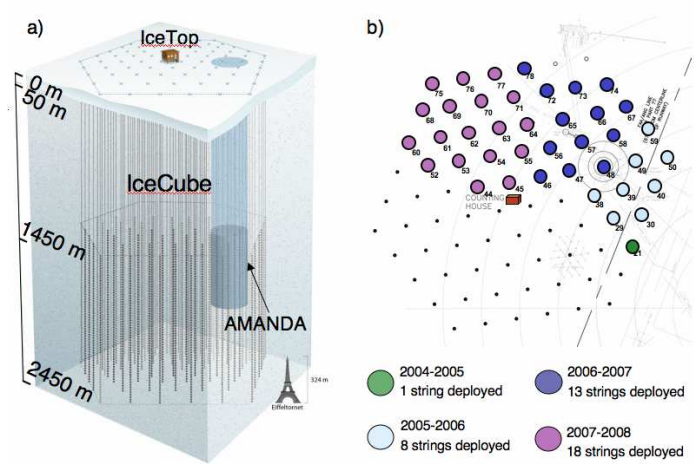}
\caption{The IceCube detector: a) side view of the complete detector with 80 strings and
b)  top view of the detector showing the construction stages. \label{fig1}}
\end{center}
\end{figure}

The IceCube detector is shown in Fig.\ref{fig1}.
When complete, the detector will cover an area of 1 km$^2$  at depths of 1.45 to 2.45 km below the surface.
The $\sim1$km$^3$ detector volume  is driven by the low extra-terrestrial neutrino flux expectations\cite{theory}  and neutrino interaction cross sections.
IceCube will be composed of $\sim4800$ in-ice Digital Optical Modules (DOMs), recording Cherenkov light from charged particles, on $80$ strings spaced by $125$ m.
%Above each string, there is a station of the surface array IceTop. Each IceTop station consists of 4 DOMs in two tanks.
The data are digitized in the DOMs and sent to the surface, where filtering algorithms are used to reduce the data volume for satellite transmission off-site.
A detailed detector description can be found in Ref.\cite{icecube-first}.

The construction of IceCube began in 2005 and proceeds in yearly stages as illustrated 
in Fig.\ref{fig1}b). Currently  $50$\% of the detector is installed and operational.
Its completion is expected by 2011.

\section{Neutrino detection}
High energy neutrinos are detected by observing the Cherenkov radiation from secondary
particles produced in neutrino interactions inside or near the detector.
Muon neutrinos from charged current (CC) interactions
are identified by the final state  muon track \cite{amanda-muon}.
Electron and tau neutrinos produced in CC interactions, as well as all neutrinos produced
in neutral current (NC)
interactions are identified by observing electromagnetic or hadronic showers (cascades).
Track and cascade reconstruction algorithms are described in detail in Refs.\cite{amanda-muon,amanda-cascade}.

The backgrounds to extra-terrestrial neutrinos of all flavors are down-going cosmic ray muons and 
atmospheric neutrinos (produced by the decay of $\pi$, $K$ and charmed mesons in cosmic ray air showers in the Earth's atmosphere)\cite{kowalski-fluxes}.  

\section{Muon neutrino results with IC9} 
\begin{figure}[t]
\includegraphics[width=0.49\textwidth]{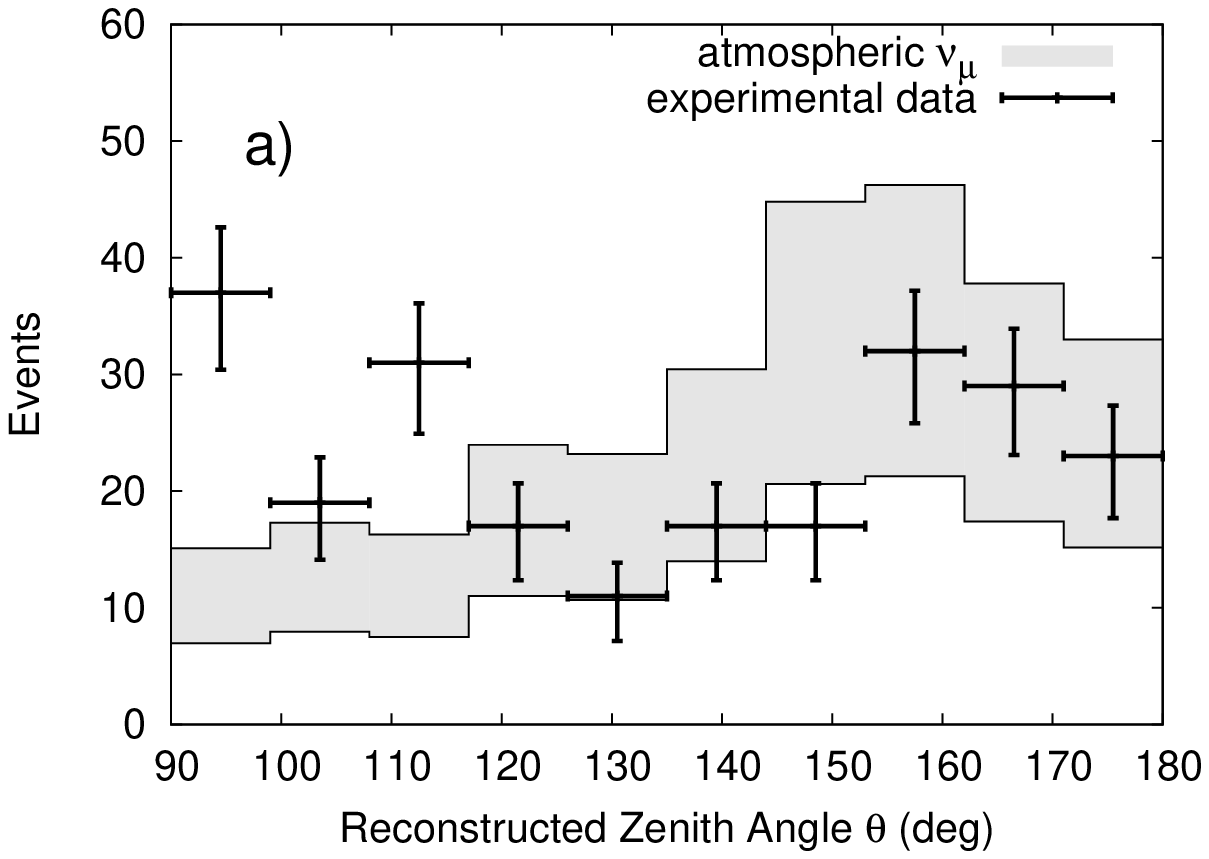}
\includegraphics[width=0.49\textwidth]{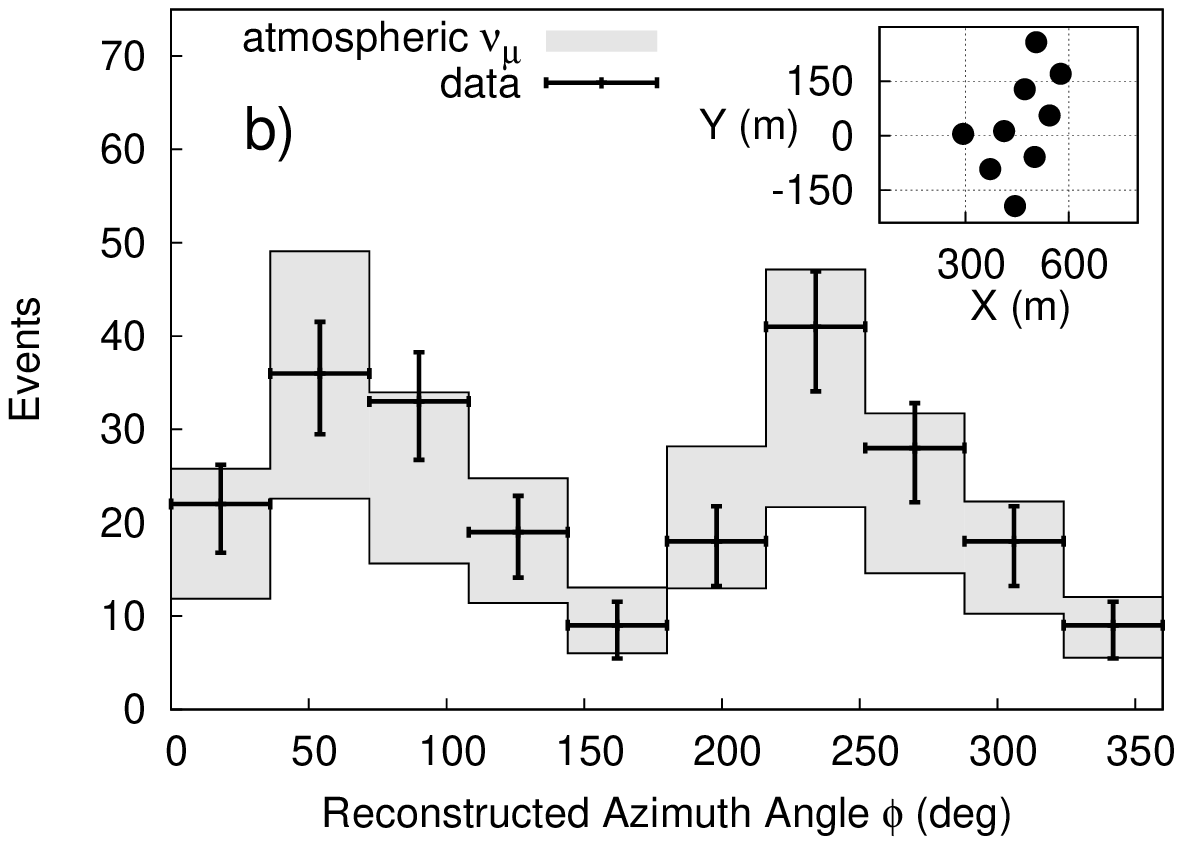}
\caption{\label{fig:zenithDistribution}
Distribution of the reconstructed:  a) zenith angle $\theta$
and b) azimuth angle $\phi$.
The error bars on the experimental data are statistical. 
The band shows the range for the atmospheric neutrino simulation. The figure is from Ref.$^{9}$.
\label{fig3}}
\end{figure}

After cuts, which selected well reconstructed up-going muon tracks, a total of $234$  
neutrino candidate events were identified in the first 137.4 days of livetime
with the nine string detector (IC9)\cite{atmospheric-9strings}.
The zenith angle distribution of those events is shown in Fig.~\ref{fig3}a).  
The small excess of events at low zenith angles (near the horizon) is most likely caused by misreconstructed down-going muons\footnote{A zenith angle of $90$ degrees indicates a horizontal event, and a zenith angle of $180$ degrees is a directly up-going event.}.
Events with larger zenith angles are  consistent with 
the simulated atmospheric $\nu_{\mu}$s. 
%The excess of data events at low zenith angles is most likely misreconstructed down-going muons,
%which are increasingly hard to reject near horizon.
Figure~\ref{fig3}b) shows the azimuth distribution of neutrino candidate events. 
The two peaks reflect the IC9 asymmetric geometry,
shown as an inset, and are well reproduced by Monte Carlo background simulations\cite{corsika}.
The agreement of the detected atmospheric neutrinos with expectations established IceCube as a neutrino telescope\cite{icecube-first,atmospheric-9strings}.  
The atmospheric muon neutrino event sample 
was used for the first extra-terrestrial neutrino point source searches with 
IceCube\cite{point-source-9strings}.
\begin{figure}[t]
\begin{center}
\includegraphics[width=0.9\textwidth]{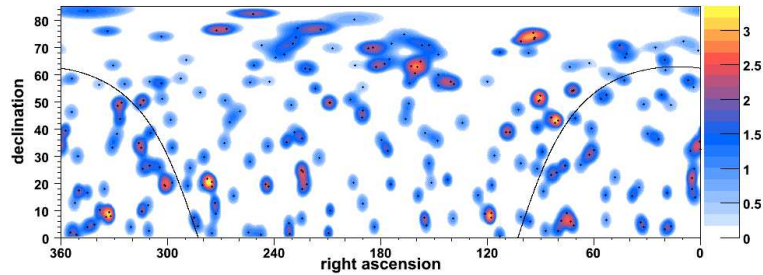}
\caption{ Sky map of the significance [$\sigma$] of deviations from background, 
estimated from the maximum likelihood point-source search.  
The figure is from Ref.$^{11}$. \label{fig4}}
\end{center}
\end{figure}
The sky-averaged point source sensitivity to a source with an $E^{-2}$ 
spectrum is $E^2 d\Phi/dE = 1.2 \cdot 10^{-7}$ GeV cm$^{-2}$  s$^{-1}$,
comparable with the limit obtained from the AMANDA-II analysis of five years of data taking
\cite{amanda-point-source}.
The results of the all-sky search are shown in Fig.\ref{fig4}. The maximum deviation from background
($3.35~\sigma$  at r.a. = $276.6\deg$ , dec= $20.4\deg$) is consistent with random fluctuations.
A search for neutrinos from $26$ galactic and extragalactic sources was also 
performed.  The most significant excess over background was $1.77\sigma$ for the Crab Nebula, consistent with  random fluctuations. 
The $90$\% C.L. flux upper limit for the Crab Nebula is 
$E^2 d\Phi/dE = 2.2 \cdot 10^{-7}$ GeV cm$^{-2}$s$^{-1}$,  Ref.\cite{point-source-9strings}. 

\section{Diffuse flux searches}
If the neutrino fluxes from individual sources are too small 
to be visible as individual sources, but  the number of sources is large,
then these neutrinos will be detectable as a diffuse flux coming from the entire sky.
The best known limit on the maximum diffuse neutrino flux
was calculated by Waxmal and Bahcall\cite{wb-limit}(WB) to be
$E^{2} d\phi/dE < 7\cdot 10^{-8}$ GeV cm$^{-2}$ s$^{-1}$ sr$^{-1}$.

\begin{figure}[t]
\begin{center}
\includegraphics[width=0.9\textwidth]{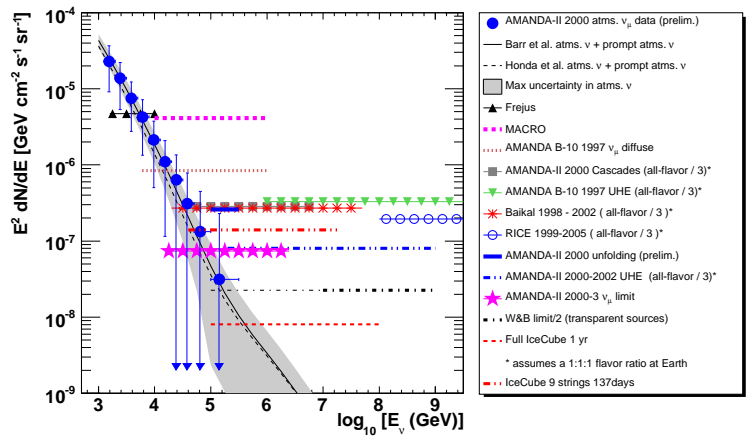}
\caption{ Summary of existing experimental limits on the diffuse neutrino flux versus the logarithm of 
neutrino energy. The figure is from Ref.$^{14}$.  \label{fig5}}
\end{center}
\end{figure}

The experimental search method assumes that the signal has a harder energy spectrum than atmospheric neutrinos. When examining energy-related  parameters, an excess of events over the expected atmospheric neutrino background would be indicative of an extraterrestrial neutrino flux. 
So far, no excess of events has been observed.  
For the IC9 data and 137 days livetime 
the expected sensitivity 
$E^{2} d\phi /dE <  1.4 \cdot10^{-7}$ GeV cm$^{-2}$ s$^{-1}$ sr$^{-1}$ 
is only a factor of $2$ larger than the tightest AMANDA-II limit\cite{kotoyo}.
The summary of  existing experimental limits on extra-terrestrial diffuse neutrino flux  and 
WB flux limits  are shown in Fig.~\ref{fig5}. 
  
\section{Summary}

The detection of atmospheric neutrinos with $9$ string configuration established IceCube as a neutrino telescope\cite{icecube-first,atmospheric-9strings}. 
A significant improvement of the sensitivity for both point-like neutrino sources 
as well as diffuse neutrino fluxes is expected from 2007 data taken with $22$ strings.
Analyses of these data are in progress. 
The IceCube detector continues to grow.  We expect  that
an integrated exposure of 1km$^3$ $\cdot$ year will be reached in 2009 
and the first extra-terrestrial neutrino signal may be detected.   
Stay tuned!  \newline

\section*{Acknowledgments}

\noindent
We acknowledge support from the U.S. National Science Foundation and Department of Energy.

\end{document}